# Beyond Words: Interjection Classification for Improved Human-Computer Interaction

Alexander Apartsin
*Holon Institute of Technology*
Holon, Israel

Yaniv Goren
*Afeka Academic College of Engineering*
Tel Aviv Israel

Yuval Cohen
*Afeka Academic College of Engineering*
Tel Aviv Israel
0000-0002-8225-6960

Yehudit Aperstein
*Afeka Academic College of Engineering*
Tel Aviv Israel
0000-0001-6390-9463

*Abstract*—In the realm of human-computer interaction, fostering a natural dialogue between humans and machines is paramount. A key, often overlooked, component of this dialogue is the use of interjections such as "mmm" and "hmm". Despite their frequent use to express agreement, hesitation, or requests for information, these interjections are typically dismissed as "non-words" by Automatic Speech Recognition (ASR) engines. Addressing this gap, we introduce a novel task dedicated to interjection classification, a pioneer in the field to our knowledge. This task is challenging due to the short duration of interjection signals and significant inter and intra-speaker variability. In this work, we present and publish a dataset of interjection signals collected specifically for interjection classification. We employ this dataset to train and evaluate a baseline deep learning model. To enhance performance, we augment the training dataset using techniques such as tempo and pitch transformation, which significantly improve classification accuracy, making models more robust. The interjection dataset, a python library for the augmentation pipeline, baseline model, and evaluation scripts are available to the research community.

## I. INTRODUCTION

People tend to choose an efficient form of verbal communication by leveraging a common understanding of context between two parties. Interjections are one of the major parts of speech frequently used to convey meaning in a specific context [1][2][3][4]. Many smart systems include a speech recognition system that enables natural dialogue between man and machine [1]. Among well-known examples are virtual personal assistants, including Apple's Siri artificial intelligence system [2], or Amazon's Alexa [3]. A Siri or Alexa user can, with natural speech queries, obtain information and perform various actions in several domains (e.g., checking weather or stock prices, ordering pizza, etc.).

In recent years, a communications revolution has expanded from a rich human-machine interaction to a full-fledged person-machine communication [5], [6], [7], [8]. This revolution requires a high-quality interface to support successful Human Computer Interactions (HCI) [8]. Adding interjection recognition capabilities to voice assistants will improve human computer interactions [9] and increase usage, with spontaneous conversation in human-machine interfaces. The improvement can be expressed by interpreting the meaning of hard-to-understand speech, such as a heavy accent, or a sentence in which not all words are clear [10]. By understanding prominent interjections, we can understand the whole sentence. For example, when a human says "Oy, …" in a conversation, he wants to express an unexpected situation, something unfortunate, or perhaps fright. Interjections often express spontaneous feelings, a key feature that can be harnessed for Speech Emotion Recognition (SER) [11].

While humans adeptly recognize emotional cues in speech, this remains a research frontier for machines. Enabling machines to comprehend and respond to emotions can enhance services like call centres, particularly in emergency situations. This emotional intelligence in machines is crucial for fostering more human-like interactions.

Voice interjections can be seen as a form of "voice touch" signal, akin to haptic touch interfaces, potentially simplifying interactions with voice assistants and other voice-enabled devices. Interjections can be mapped to specific actions or responses, much like shortcuts in a user interface. For instance, positive interjections like "aha" could affirm a process's continuation, while "uh-oh" could signal a need for probing if something went wrong. This allows for the replacement of lengthy commands with simple interjections, customizing the system to individual user preferences and needs. For example, a user could choose an interjection to represent a frequently used command, streamlining their interaction with the system.

It is important to emphasize that our aim is not to detect specific interjection phrases within speech or to build a system that recognizes all interjections. Instead, our research presents a benchmark system for interjection classification, providing a solid foundation for the quick and simple addition of new interjections. Our motivation is to enhance user interface technologies, enabling the creation of more natural and habitable human-machine interfaces and dialogues.

The collection and the preparation of the training data is a major challenge for this project. The available speech datasets are focused on word level [12], phoneme level [13], or event level tasks [14], while interjections lie somewhere in-between. Therefore, we collected our own unique baseline dataset by recording a relatively large set of interjections and negative examples from some speakers.

In this work, we propose a neural network model for interjection classification. Since deep learning models require a large amount of training data, we enriched the dataset by augmenting the original data through the application of various artificial distortions [15]. The da-ta augmentation includes the addition of background noise [16] and pitch and tempo modifications [17].

In this paper, we present results of several different interjection classification baseline experiments, where relative improvement obtained by using the proposed data augmentation methods. By incorporating these augmentation methods into the training process, we enhanced the network's ability to handle variations, thereby improving its capacity to generalize and perform effectively on new, unseen data.

The rest of this paper is organized as follows: We begin with a review of prior work, focusing on speech recognition, augmentation methods in this field, and research related to interjections. Subsequently, we provide an in-depth explanation of this research contribution. This is followed by an overview of the dataset creation process and a description of our data augmentation tool. We then introduce the interjection classification models and delve into the specifics of our experimental setup, which is designed to evaluate the effectiveness of our approach in various scenarios. Finally, we discuss the results, summarize the proposed method, and suggest potential avenues for future research.

## II. PREVIOUS WORK

One of the most common tasks associated with processing speech and voice signals is Automatic Speech Recognition (ASR) [18]. ASR is the task of translating audio signals into text. The main challenge is to overcome the non-stationarity of the speech signal and the large variations in its spatio-temporal representation. The typical ASR system usually includes several steps. The pre-processing step, where an analog speech signal is trans-formed into a digital signal, includes speech/non-speech segmentation and filtering. The feature extraction step deals with the transformation of the incoming digital waveform into a vector representation of desirable speech features that emphasize linguistic information. Usually, a speech signal is broken into short (usually around 20-30 ms) segments (frames), overlapped every 10 ms, during which the signal is assumed to be stationary [19], [20]. During the last few decades, several techniques have been developed for feature extraction from speech signals. These approaches include Mel-Frequency Cepstral Coefficients (MFCCs), Perceptual Linear prediction (PLP), Relative Spectral (RASTA), and Linear Predictive Coding (LPC) [21]. However, MFCCs are probably used most commonly used [22].

MFCCs extract spectral features by defining an analysis window (around 25 ms) and divide the speech signal into different time frames by shifting the window with a 10 ms shifting stride. Then Fast Fourier Transformation (FFT) is calculated for each frame to obtain the frequency features, and the logarithmic Mel-Scaled filter bank is applied to its power spectrum estimate. MFCCs calculate the Discrete Cosine Transformation (DCT) of the log energies in the corresponding frequency bands to obtain an m-dimensional coefficients vector. The measured power spectrum envelope in each frame correlates to the shape of the vocal tract, providing an appropriate representation of the sound or phoneme being produced. This procedure results in feature vectors that can be arranged in a [n×m] matrix, where m is the number of coefficients and n is the number of frames.

At the heart of an ASR system is the decoder. During this phase, feature vectors are decoded into linguistic units that make up speech. The decoding relies on acoustic and language models [23] to recover the most probable utterance by modelling the conditional probability of the nth word, using the (n-1) earlier words. Linguistic and pronunciation dictionaries are often used to improve the decoding performance. An acoustic model [23] is a fundamental part of the ASR system, where the connection between the acoustic information (e.g., feature vectors) and phonetics is established. The acoustic models are frequently implemented using various methods that include the Hidden Markov Model (HMM) [24] and support vector machines (SVM) [25]. HMM is the most commonly used acoustic model for speech recognition in many practical applications.

ASR is a broad topic that includes subtopics related to interjection recognition. One of those subtopics is Key-word Spotting (KWS) [26]. The need to provide users with a fully hands-free experience for situations like driving resulted in the development of a system that listens continuously for specific keywords to initiate voice input. Keyword Spotting aims at detecting predefined keywords in an audio stream. A commonly used technique for keyword spotting is the Keyword/Filler Hidden Markov Model (HMM). One disadvantage with this technique is that it can be computationally expensive, and the model is trained separately for each keyword [27].

Keyword spotting is sometimes performed using a pattern matching approach where the input is compared with a few prerecorded commands. Since the same word might be articulated with a different speed, the Dynamic Time Warping (DTW) algorithm is used to align two sequences in an optimal way [28]. Recent neural network models show a significant improvement over the HMM approach. The recurrent neural networks (RNN) used in the Deep KWS model. [27] shows good performance while keeping a reduced runtime computation and smaller memory footprint. A Convolutional Neural Network (CNN) architecture de-scribed by [29] shows improvements over FNN in a variety of small and large vocabulary tasks. The described CNN architecture generalizes more easily to different speaking styles compared to a fully connected FNN architecture.

Another sub-topic related to interjection recognition is Audio Event Detection (AED) [14], which is considered a common task of processing speech and voice signals. An audio event is a specific type of sound, such as footsteps, running water, exhaust fan noise, screams, ocean waves breaking, or music. Many sound clips contain multiple acoustic events that overlap on the time axis. AED is a task in which a relatively long (several seconds to tens of seconds) sound clip including multiple acoustic events serves as input, and the output is acoustic event labels and their time stamps (start and end times). The process of AED is created by extracting acoustic features using MFCC in general, and then constructing classification models using HMM [24], SVM [25] or more recently models such as CNN [29] or RNN [25]. Because acoustic events can often have a temporal overlap, the most difficult problem in AED is how to detect active durations of acoustic events. For instance, Mulimani et al. [31] proposed acoustic event detection based on a convolutional neural network, which calculates a posterior for the existence of acoustic events time frame by time frame.

Generally, deep learning requires a large amount of labelled training data to enable accurate speech recognition. To the best of our knowledge, such a large dataset does not exist for interjections. Therefore, data augmentation is proposed, where the speech data are artificially augmented by applying different types of distortions in a way that does not change the label. Stoian et al. [18] and Singh et al. [32] proposed data augmentation for low resource speech recognition tasks. The performance of an FNN speech recognition model depends on how well the training data matches the testing data. This can be done by increasing the generalization of a model be-yond the data provided to it, and by attempting to create similarity between the training data and the representative characteristics that are seen in real data, such as voice variability of different speakers, or different back-ground noises.

There are many options for data augmentation. Some of them are applied in the feature level of a neural network, and some are applied directly in the raw audio level. Vocal tract length perturbation (VTLP) [33] is a popular method for doing feature level data augmentation in speech and has shown gains on the TIMIT (a collection of phonemically transcribed American English speech) phoneme recognition task. SpecAugment [34] is another feature level data augmentation method for speech recognition that operates on the log Mel spectrogram of the input audio. In the raw audio level, intuitive and practical transformations, such as Dynamic Range Compression (DRC), pitch-shifting, time stretching, and background noise combination apply audio effects to the original training audio files [35].

Despite the prevalence of interjections in human speech patterns, interjections in literature are mentioned mainly in the context of emotion, where researchers seek to understand the nature of interjections and interjectional meaning [36],[37]. There is a difference between formal speech and conversational speech. Conversational speech is more spontaneous and efficient and does not require special training. Székely [38] describes fundamental properties of spontaneous speech that present problems for spoken language applications, such as lack of punctuation or the inability to "hear" a speaker's emotion through speech. As stated before, SER is still an ongoing subject of research and the main traditional techniques for SER are based on feature ex-traction and selection to identify various emotions [11].

Although interjections have been acknowledged for their emotional and communicative functions, their integration into speech technologies remains limited. Cohn et al. [39] explored the impact of adding interjections as emotional-cognitive cues in text-to-speech systems, demonstrating improvements in perceived conversational quality. Similarly, Yoon et al. [40] investigated fear emotion classification in speech using interjection-related cues, treating them as indicative features of heightened emotional states. However, these studies did not address interjections as standalone acoustic events or attempt to classify their types explicitly. Instead, interjections were used indirectly, serving broader affective or stylistic functions within spoken interactions.

III. OUR CONTRIBUTIONS

In this work, we propose a novel approach to the task of interjection classification, employing a fully connected feedforward neural network (FNN). The task presents unique challenges, given the nature of interjections as short, context-free sound fragments that exhibit significant variability within and between speakers.

To our knowledge, no existing dataset is specifically tailored to the study of interjections. As such, we have taken the initiative to create a new dataset comprised of interjection phrases from various speakers. The interjection categories were selected for their language-independence and semantic clarity, enabling consistent modeling across different speakers. This dataset serves as a valuable resource for research in this area.

While previous studies have considered interjections primarily as paralinguistic or affective cues within broader emotion recognition tasks, we treat interjections as the central classification targets. Furthermore, we introduce the use of audio data augmentation as a strategy to mitigate the limitations posed by data scarcity. We investigate the impact of different augmentation techniques on the performance of our proposed architecture. By training the network on augmented data, we aim to enhance the network's robustness to deformations and improve its generalizability to unseen data in both quiet and high-noise environments.

IV. METHODOLOGY

We aim to create a baseline machine learning model for interjection classification. This model will process audio input and assign labels, distinguishing between interjections and non-interjection words. To train this model, we created a 'clean' dataset from various speakers. Initially, we use data from a single speaker to focus on interjection recognition. The dataset is later expanded with more speakers and augmented with artificial modifications to pitch, tempo, and added background noise.

A. Construction of Original Dataset

Our dataset consists of 'clean', unsynthesized audio samples spanning five distinct categories: four dedicated to different interjections and one to non-interjection words. The latter category encapsulates words randomly selected from a book. The interjection categories were specifically chosen due to their language-independence and their ability to convey inherent semantic meaning, irrespective of conversational context. Table I presents the selected interjections.

TABLE I
FOUR SELECTED INTERJECTIONS THAT RECORDED
INTO SEPARATE SOUND CLIPS

| interjection | alternate/similar | translation/meaning |
|---|---|---|
| nah | | **"No"** - Informal no |
| mmm | mhm, uh-hu | **"Yes"** - Agreement, acknowledgement |
| ahah | Aha, ahh | **"I understand"** - Understanding, Confirmation |
| oy | oy vey | **"Oh no..."** - Mainly Jewish: Used to express self-pity, or expression of unexpected situation |

To serve the purposes of this research, we recorded five speakers, comprising two females and three males, each contributing a certain number of audio samples for every category. The column "Number of Audio Samples" refers to a specific word, e.g., Speaker A has 850 audio samples per word and a total number of 4,250 audio samples. The speaker profiles and the count of audio samples contributed by each speaker for every category are detailed in Table II. A profile denoted as a triple (X, Y, Z) signifies X as the gender (M for male, F for female), Y as the age, and Z as the speaker's mother tongue (H for Hebrew, S for Spanish).

TABLE II
SPEAKER PROFILES AND NUMBER OF AUDIO
SAMPLES RECORDED FOR EACH SPEAKER AND
WORD

| SPEAKER | NUMBER OF AUDIO SAMPLES | PROFILE |
|---|---|---|
| A | ~850 | M, 44, H |
| B | ~550 | F, 42 H |
| C | ~300 | M, 16, H |
| D | ~300 | F, 81, S |
| E | ~300 | M, 50, H |

The sound clips were recorded using Auditok, which is a Voice Activity Detection tool [41] that enables recording and saving each sound clip as a separate ".wav" file with 16000 Hz sampled rate.

*B. Data Preprocessing*

All audio files in our dataset were standardized to a uniform length. This was done to ensure a consistent number of frames across all files during the feature extraction phase, resulting in fixed-length vectors for our classification model. We chose a length of 1.55 seconds, which accommodates longer interjection terms. We also set a minimum recording length of 0.45 seconds to exclude short, unrelated recordings such as background noises, and a maximum length of 1.55 seconds to account for instances where Auditok did not detect silence between recordings. Any recordings falling outside this length range were excluded from the dataset.

To achieve a fixed length, silence was added to the beginning and end of each shorter sound. This ensures that the actual sound portion of each recording is unevenly distributed across the fixed length. For instance, one recording might begin with 0.2 seconds of silence and end with 0.74 seconds of silence, while another of the same label might start with 0.61 seconds of silence and conclude with 0.3 seconds of silence. This approach prevents the model from learning the exact onset of the word's sound. Following this process, we generated five folders, corresponding to four interjections and one negative example, each containing approximately 800 fixed-length recordings.

To extract the useful features from the audio file, Librosa library [42, 43] has been used. This library is a python package for music and audio analysis that provides several methods to retrieve information from sound clips. The methods from Librosa we have been using to extract various features are: MFCC, Melspectrogram, Chorma-stft, Spectral contrast and Tonnetz. The result of that process is a matrix with a row for each audio file, and a column for each mean feature value (193 columns).

*C. Data Augmentation System (DAS)*

To extend the collected dataset, a data augmentation system (DAS) that generates synthetic samples was developed. We used Pysox [44, 45], a Python library that provides a simple interface between Python and Sox, a popular command line tool for sound processing that can apply various effects to audio files. We tried four different sound effects on our original unsynthesized ("clean data") audio samples. Each effect was applied directly to the sound file prior to converting it into the input vector representation for the neural network. The data augmentation effects are described below:

• *Tempo*: Change the audio playback speed but not its pitch. This function gets the parameter 'factor' (factor > 1 speed up the audio signal, factor < 1 slow down the audio signal). The duration of the sound file is changing.

• *Pitch*: Change the audio pitch (but not tempo). The sensation of a frequency is commonly referred to as the pitch of a sound. A high pitch sound corresponds to a high frequency sound wave and a low pitch sound corresponds to a low frequency sound wave. One octave (the interval between one musical pitch and another) is divided into 12 semitones (tones) of 100 cents each. Typically, cents are used to express small intervals. 1200 cents equal to 1 octave. This function gets the parameter 'n-semitones' (The positive or negative number of semitones to shift). Duration of the sound file is not changing.

• *Background noise (BGN)*: Mixes the original "clean" audio with another recording containing background sounds from several acoustic scenes. Each sample is mixed with nine acoustic scenes: baby gibberish, ambulance, crowd laughing, football crowd, mall, passing bus, rain, street traffic, and TV. Each mix $S_{mix}$ was generated using (1)

$$S_{mix} = (W_{orig} \times S_{orig}) + (W_{BGN} \times S_{BGN}) \quad (1)$$

where $S_{orig}$ is the original audio sample and $S_{BGN}$ is the signal of the background scene. $W_{orig}$ and $W_{BGN}$ are weighting volume parameters determined such that $W_{orig} + W_{BGN} = 1$. For each $S_{orig}$ audio, two or three $S_{mix}$ audio files with different weights were generated, depending on the dataset.

DAS can create a large number of synthetic samples from original "clean" samples. The number of synthetic samples is determined by the configuration of the methods to use (tempo, pitch, and background noise are the three supported methods now, but any other method can be easily added), and by configuration of each method value. For example, if the configuration for the pitch method contains four different values for the semi-tone parameter, then the system will generate four different augmented audio samples with the desired pitch level.

A significant advantage in DAS is the fact that different effects can be combined for each recording. This allows us to control the size and the diversity of the dataset we want to create; e.g., if the desired methods are tempo (with factors 0.9 and 1.1) and pitch (with semitones 2 and -2), an additional eight audio samples will be created from one original audio (two audios with only tempo effect, two with only pitch, and four that are result of combining tempo and pitch parameters together).

In addition, DAS can generate white noise in the background of each original audio file. This functionality was added because white noise is produced by combining sounds of all different frequencies together and can be used to mask other sounds, such as background noises.

### D. Network architecture and Evaluation metric

In our research, we employ a fully connected Feedforward Neural Network (FNN) to classify single-word audio files into one of five categories, four of which are interjections and one non-interjection. This architecture, implemented via TensorFlow, is tailored to the multiclass nature of our task. The softmax activation function in the output layer generates a probability distribution vector of potential outcomes, facilitating the classification process.

To measure the accuracy of our model, we utilize the F1 score, which balances precision and recall. This metric is particularly suitable for our task as it accounts for both false positives and false negatives. Therefore, our model is not only designed to accurately identify interjections but also to minimize misclassifications, a critical aspect for enhancing human-computer interaction.

## V. EXPERIMENTAL SETUP

Our experimental setup is designed to evaluate the effectiveness of our approach in various scenarios. We first present augmented datasets using the Data Augmentation System (DAS). We then present two main evaluation scenarios devised to examine the impact of data augmentation on the effectiveness of interjection classification in both quiet and high-noise environments.

We have made both the source code and the interjections dataset available to the research community: https://github.com/aperstein/interjection

### A. Construction of Augmented Datasets using DAS

It is important to choose the augmentation parameters such that the semantic validity of the label is maintained. To create our augmented datasets, we chose several parameters: for tempo, a factor in the range of 0.86-1.14; for pitch, a semitone in the range of (-2.4)-2.4; and for background noises, $W_{orig}$ in the range of 0.83-0.93 and $W_{BGN}$ in the range of 0.07-0.17, respectively In our research, we developed augmented datasets by applying tempo, pitch, and background noise effects, both individually, and in various combinations. The resulting augmented sets are presented in Table III.

TABLE III
DESCRIPTION OF THE AUGMENTED DATASETS

| | ORIGINAL SAMPLES PER CLASS | METHODS USED | GENERATED SAMPLES PER CLASS |
|---|---|---|---|
| 1 | 120 | Pitch | 2520 |
| 2 | 120 | Tempo | 2520 |
| 3 | 100 | BGN | 2900 |
| 4 | 60 | Tempo + Pitch | 4860 |
| 5 | 40 | Pitch + BGN | 6880 |
| 6 | 40 | Tempo + BGN | 6880 |
| 7 | 10 | Tempo + Pitch + BGN | 9320 |

For each speaker, 120 different audio samples were used from our original dataset (600 files altogether from four interjection folders and one folder of negative examples), whose overall length is about eight minutes; to generate seven augmentation sets with a total of 180,000 audio samples, whose overall length was about 78 hours. For all speakers, the system generated more than 300 hours of augmentation data from 32 minutes of original recordings.

*B. Evaluation Scenarios*

Our study is structured around two primary scenarios. The first scenario is designed to assess how data augmentation can enhance the performance of interjection classification for speakers not previously encountered by the model. We initiate this scenario by training the classifier using clean data from two speakers (one male and one female). The model is then tested on clean data from two different speakers (also one male and one female). The data from the remaining subject is utilized as a validation set for optimizing hyperparameters. Subsequently, the classifier is trained separately on each of our seven augmented datasets and retested on the clean data from the unseen speakers.

In the second scenario, we explore the impact of data augmentation on classification performance in real-world, high-noise environments. We posit that a system trained with augmented data can effectively handle real external sound sources. The initial phase of this scenario involves training the classifier using clean data from a single speaker. The model is then tested on data derived from audio signals recorded from the same speaker in a real-world noisy environment, which includes background noises such as television playing, children's sounds, a baby crying, and an air conditioner running. In the subsequent phase, the classifier is trained using our artificially augmented data sets from the same speaker. The model's accuracy is then tested against the same natural background data set used in the first phase of this scenario.

The models were trained using the Adam optimizer, with a learning rate of 0.009 and 3 hidden layers. These parameters were selected after extensive experimentation with various alternatives. For instance, we tried training epochs ranging from 150 to 300, adjusted the learning rate starting from 0.1 and reducing it for better convergence, experimented with different activation functions, number of hidden layers and tested different optimizer algorithms like classic SGD and Adam.

## VI. RESULTS AND DISCUSSION

Two experiments were conducted in our study. In both the accuracy assessment was measured by comparing the baseline (original dataset) accuracy with the accuracy level of each of the proposed augmentation methods.

The results of the first experiment are presented in Table IV. The model trained with samples of two subjects (female speaker A and male speaker B) and tested separately on previously unseen speakers (male unseen speaker C and female unseen speaker D). The data from subject E was utilized as a validation set for optimizing hyperparameters. The percentages within the parentheses present the accuracy improvement compared the baseline clean dataset accuracy. We can realize that each of the augmented methods significantly improves accuracy relative to the baseline accuracy. Each of the augmented sets were helpful in this scenario, and the highest classification accuracy improvement for each unseen speaker was achieved by the dataset combined with all three augmentation methods, where an increase of 63.6% was obtained for speaker C and 42.9% was obtained for speaker D, which confirms our claim that the combination of the augmentation methods gives better results.

TABLE IV
RESULTS OF THE FIRST SCENARIO

| TRAINING DATASET (SPEAKERS A+B) | ACCURACY (SPEAKER C) | ACCURACY (SPEAKER D) |
|---|---|---|
| Original clean data | 0.286 | 0.41 |
| Pitch | 0.391 (36.7%) | 0.44 (7.3%) |
| Tempo | 0.405 (41.6%) | 0.477 (16.3%) |
| BGN | 0.418 (46.2%) | 0.415 (1.2%) |
| Tempo + Pitch | 0.456 (59.4%) | 0.438 (6.8%) |
| Pitch + BGN | 0.454 (58.7%) | 0.484 (18%) |
| Tempo + BGN | 0.393 (37.4%) | 0.529 (29%) |
| Tempo + Pitch + BGN | 0.468 (63.6%) | 0.586 (42.9%) |

The results of the second experiment are presented in Table V, where the impact of augmentation in natural high noise environment was tested.

TABLE V
RESULTS OF THE SECOND SCENARIO.

| TRAINING DATASET | ACCURACY |
|---|---|
| Original clean data | 0.45 |
| Pitch | 0.344 (-23.6%) |
| Tempo | 0.476 (5.8%) |
| BGN | 0.63 (40%) |
| Tempo + Pitch | 0.424 (-5.8%) |
| Pitch + BGN | 0.569 (26.4%) |
| Tempo + BGN | 0.667 (48.2%) |
| Tempo + Pitch + BGN | 0.567 (26%) |

After training the classifier with clean data of one speaker and testing it with natural background noises, we trained it with the augmented datasets we created for the same speaker. We average accuracy over all subjects. Highest classification accuracy improvement achieved by combining tempo and artificial background noises, where in several of the testing the accuracy was beyond 0.7. However, in contrast to the strong impact artificial background noises, and the slight positive effect of the tempo, the impact of the pitch was negative. A better future experiment will be to try different semitones values for this augmentation method.

In both scenarios the model seems to be more robust when using the background noise method. In most cases (6 from 9 cases) the combination between background noises and other methods improved the accuracy of the classifier in relation to use only the background synthesis. It implies that using some different settings for each effect can improve the results even more than the results we obtained in the scenarios described.

## VII. Conclusion

In this research, we have ventured into the largely unexplored territory of interjection classification in human-computer interaction. Since interjections are important in fostering natural dialogue [46], but are often dismissed as "non-words", we have initiated this research dedicated to their classification. This research, however, is quite challenging, given the short duration of interjection signals and the significant variability within and between speakers.

To tackle these challenges, we created a unique dataset of interjection phrases from various speakers and have developed a model for interjections classification. The constructed dataset, the first of its kind, serves as a valuable resource for further research in this area.

Recognizing the limitations posed by data scarcity, we have also introduced the use of audio data augmentation. By training our network on augmented data, we have enhanced its robustness to deformations and improved its generalizability to unseen data. This approach has proven particularly effective in both quiet and natural high-noise environments.

Our findings underscore the potential of our approach to enhance human-computer interaction by incorporating interjections, paving the way for more natural and intuitive dialogues that truly reflect the nuances of human communication.